# Low-Temperature Magnetic Susceptibility of $ErAl_3(BO_3)_4$ Single Crystal: Evidence of Antiferromagnetic Spin-Spin Correlations

D. N. Merenkov, V. A. Bedarev, S. N. Poperezhai, Yu. A. Savina

*B. Verkin Institute for Low Temperature Physics and Engineering of the National Academy of Sciences of Ukraine, Nauky Ave. 47, UA-61103, Kharkiv, Ukraine*

T. Zajarniuk, A. Szewczyk

*Institute of Physics, Polish Academy of Sciences,*
*Al. Lotnikow 32/46, Pl-02668, Warsaw, Poland*

I. A. Gudim

*L. V. Kirensky Institute of Physics, Federal Research Center KSC, Siberian Branch of the Russian Academy of Sciences, 660036, Krasnoyarsk, Russia*

e-mail: dmnime@gmail.com

ABSTRACT

Temperature dependence of magnetic susceptibility of the $ErAl_3(BO_3)_4$ single crystal was measured in the temperature range of $T = 2 - 15$ K at the external magnetic field of $H = 100$ Oe applied along the *a*- and *c*- crystallographic axes. The influence of spin-spin correlations on the magnetic susceptibility has been identified. It was estimated that the value of Curie-Weiss temperature is $\theta_{AF} \approx - 0.45$ K.

АНОТАЦІЯ

до статті В.А. Бєдарєв, Д.М. Меренков, С.М. Попережай, Ю.О. Савіна, T. Zajarniuk, A. Szewczyk, І.А. Гудім

«Низькотемпературна магнітна сприйнятливість монокристалу $ErAl_3(BO_3)_4$: прояв антиферомагнітних спін-спінових кореляцій»

Температурну залежність магнітної сприйнятливості монокристала $ErAl_3(BO_3)_4$ виміряно в інтервалі температур $T = 2 - 15$ K у зовнішньому магнітному полі $H = 100$ Е уздовж кристалографічних осей *a* і *c*. Виявлено вплив спін-спінових кореляцій на магнітну сприйнятливість. Проведено оцінку величини температури Кюрі-Вейса $\theta_{AF} \approx - 0.45$ K.

# INTRODUCTION

Compound $ErAl_3(BO_3)_4$ belongs to rare-earth aluminum borates, which reveal pronounced nonlinear [1] and luminescent [2] optical properties. The giant magnetoelectric effect [3] has been recently discovered in these compounds, which are also reported as promising materials for magnetic cooling [4]. Representatives of the rare-earth aluminum borate family are usually have the very low temperature of magnetic ordering and at least in one of them - $TbAl_3(BO_3)_4$ – the phase transition to a magneto-ordered state at $T = 0.68$ K is observed [5].

Similar to other rare-earth aluminum borates, the crystal structure of $ErAl_3(BO_3)_4$ has a trigonal symmetry with space group $R32$ ($D_3^7$) without an inversion center. The values of the lattice constants at room temperature are: $a = 9.2833(7)$ Å, $c = 7.2234(6)$ Å [6]. Above 2 K $ErAl_3(BO_3)_4$ is in the paramagnetic state [7]. The ground multiplet $^4I_{15/2}$ ($S = 3/2$, $L = 6$, $J = 15/2$) of the $Er^{3+}$ ions in $ErAl_3(BO_3)_4$ is splitted by the crystal field into 8 Kramers doublets. The first excited and the second excited doublets are separated from the ground doublet by the energy interval $E_1 = 66$ K (46 cm$^{-1}$) and 150 K (104 cm$^{-1}$), respectively [8].

The electron paramagnetic resonance (EPR) absorption spectra of the $ErAl_3(BO_3)_4$ single crystal were previously measured at $T = 4.2$ K in the magnetic fields applied along the $c$-axis and $a$-axis. In addition to the intensive main resonance line, the additional line with much weaker intensity is observed in the EPR spectra for $\mathbf{H} \parallel \mathbf{a}$ and $\mathbf{H} \parallel \mathbf{c}$ [9]. The obtained value of the $g$-factor of the additional line was almost two times larger than the $g$-factor of the main line for both orientations of the external magnetic field. The additional excitation had a non-zero splitting at $H = 0$. Further, the transformation of the main line into the triplet structure was revealed, as well as the presence of a minimum at $T = 4.5$ K on the temperature dependence of the heat capacity of $ErAl_3(BO_3)_4$ was observed [10]. These experimental data indicated the occurrence of a short-range magnetic order in $ErAl_3(BO_3)_4$ at liquid helium temperatures. The calculations conducted according to the suggested model of paired clusters, which takes account of the anisotropy of spin-spin interaction, well corresponded to the data from the experimental measurements. Estimated values of the spin-spin interactions are of $0.1 - 1$ K, which is in agreement with the order of magnitude for dipole-dipole interactions between the $Er^{3+}$ ions in the studied compound.

As it has been previously reported based on the measurements of magnetic susceptibility $\chi$ of $ErAl_3(BO_3)_4$ crystal at $H = 1$ kOe [7], this compound manifests well-pronounced easy plane anisotropy at low temperatures. At $T = 5$ K, magnetic susceptibility measured along the $c$-axis is less than 5% of its value measured in the direction of the $a$-axis. Such magnetic anisotropy is due to the influence of the crystal field on the electronic structure of a rare-earth ion in paramagnets of this family.

As mentioned above, the ground state of the erbium ion is a Kramers doublet, while the first excited doublet is located from it by $E_1 = 66$ K. Because of this, the magnetic susceptibility of the non-interacting $Er^{3+}$ ions in a crystal at considerably low temperatures and at considerably small fields $g_0\mu_B H \ll kT$ ($\mu_B$ is the Bohr magneton, $k$ is the Boltzmann constant, $g_0$ is the value of the $g$-factor for the ground doublet) is mainly determined by the ground doublet of the multiplet $^4I_{15/2}$. It is known that under such conditions, the temperature dependence of magnetic susceptibility is as follows: $\chi(T) \sim 1/T$ (well-known as a Curie law for ensemble of non-interacting magnetic particles). The departure of the experimental data from this law can indicate the presence of spin-spin correlations and/or Van Vleck paramagnetism. The purpose of this paper is to determine the influence of the spin correlations on the magnetic susceptibility of the $ErAl_3(BO_3)_4$ single crystal.

In order to achieve the purpose of the research, the low temperature magnetic susceptibility of $ErAl_3(BO_3)_4$ in the temperature range from 2 to 15 K, in magnetic field applied along the $a$- and $c$-axes were measured. This experiment allows to compare the influence of the spin-spin interactions on $\chi_a(T)$ and $\chi_c(T)$. So the results of such measurements can give a more precise and detailed understanding of the nature of spin-spin interaction compared to the earlier research [9, 10].

## EXPERIMENT

The single crystal of $ErAl_3(BO_3)_4$ was grown in the Institute of Physics of the Siberian Branch of the Russian Academy of Sciences in Krasnoyarsk by using a solution-melt method [11]. The orientation of the crystallographic axes was determined using the X-ray method.

Magnetic susceptibility measurements were carried out in temperature interval from 2 to 15 K by using the magnetometer PPMS-9T ("Quantum Design"). The chosen value of the external magnetic field $H = 100$ Oe well satisfied the condition $g_0\mu_B H \ll kT$ for the entire temperature range of measurement. In order to obtain the temperature dependence of molar magnetic susceptibility of the $ErAl_3(BO_3)_4$ single crystal, the measured magnetic moment of the sample was normalized to the mole ($m/M_\mu$) (where $m = 40$ mg is sample mass, $M_\mu$ is the molar mass of $ErAl_3(BO_3)_4$) and to the magnitude of the magnetic field $H = 100$ Oe, in which the measurements were made. The accuracy of setting and controlling the sample temperature was better than 0.05 K. The absolute error of the measured magnetic moment did not exceed 1 %.

Temperature dependence of molar magnetic susceptibility of the $ErAl_3(BO_3)_4$ single crystal with mass $m = 40$ mg measured at $H = 100$ Oe for two orientations $\mathbf{H} \parallel \mathbf{a}$ and $\mathbf{H} \parallel \mathbf{c}$ is shown in Fig. 1. As one see in Fig. 1, the value of $\chi_a$ significantly (more than two orders of magnitude) exceeds $\chi_c$ in the whole temperature range of investigation. This is generally in agreement with the corresponding magnetic data obtained earlier in [7].

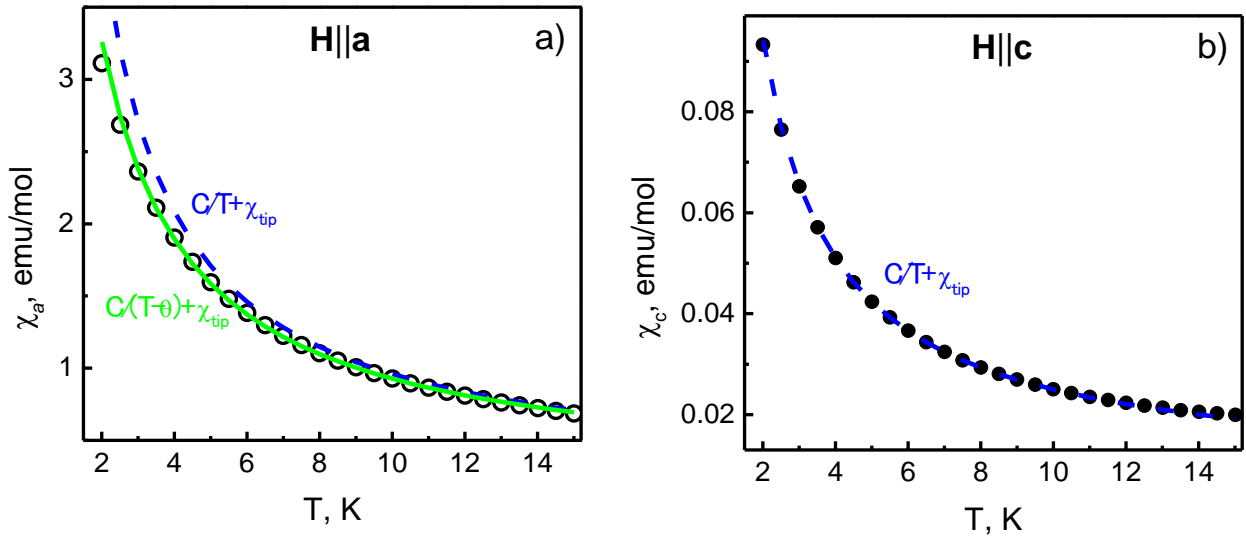

Fig. 1. Temperature dependence of molar magnetic susceptibility of $ErAl_3(BO_3)_4$ single crystal measured at $H = 100$ Oe for two orientations $\mathbf{H} \parallel \mathbf{a}$ (a) and $\mathbf{H} \parallel \mathbf{c}$ (b). Symbols are the experimental data. Solid and dashed lines show the calculation results by using expressions (4) and (1), respectively. All magnetic parameters are listed in Table 1.

For two orientations $\mathbf{H} \parallel \mathbf{a}$ and $\mathbf{H} \parallel \mathbf{c}$, monotonic increase of magnetic susceptibility was observed during the sample cooling without any manifestations of peculiarities.

## DISCUSSION

The analysis of the experimental data showed that the measured temperature dependences can not be described by the simple function $\chi(T) \sim 1/T$, even at the highest temperatures of the studied temperature range. The initial attempts to take into account spin-spin interactions did not give any positive results too. It was obvious, however, that there is at least one more factor influencing magnetic susceptibility. Following these considerations, we have replotted the molar magnetic data of our measurements as $\chi T$ versus $T$.

Temperature dependence of the product $\chi T$ for the $ErAl_3(BO_3)_4$ single crystal for two orientations $\mathbf{H} \parallel \mathbf{a}$ and $\mathbf{H} \parallel \mathbf{c}$ is shown in Fig. 2. The corresponding curves are presented as symbols. Their comparison shows one fundamental similarity – in both cases the temperature interval with a linear dependence of the product $\chi T$ and with a nonzero slope is observed. Two insets show in detail a narrow temperature interval from 6 to 9.5 K, in which a linear regime of product $\chi T$ for two magnetic field orientations is observed.

Above 9.5 K the experimental data for both orientations deviate from the linear regime. While for the case of $\mathbf{H}\parallel\mathbf{a}$, the experimental points move downwards, for the case of $\mathbf{H}\parallel\mathbf{c}$ they depart upwards. However, this is not the single difference. On cooling the sample below 6 K, the temperature dependence of the product $\chi_a T$ significantly departs from the linearity and its slope increases with lowering temperature. At the same time, for $\mathbf{H}\parallel\mathbf{c}$ $\chi_c T$ keeps the linear regime and the slope practically does not change below 9.5 K.

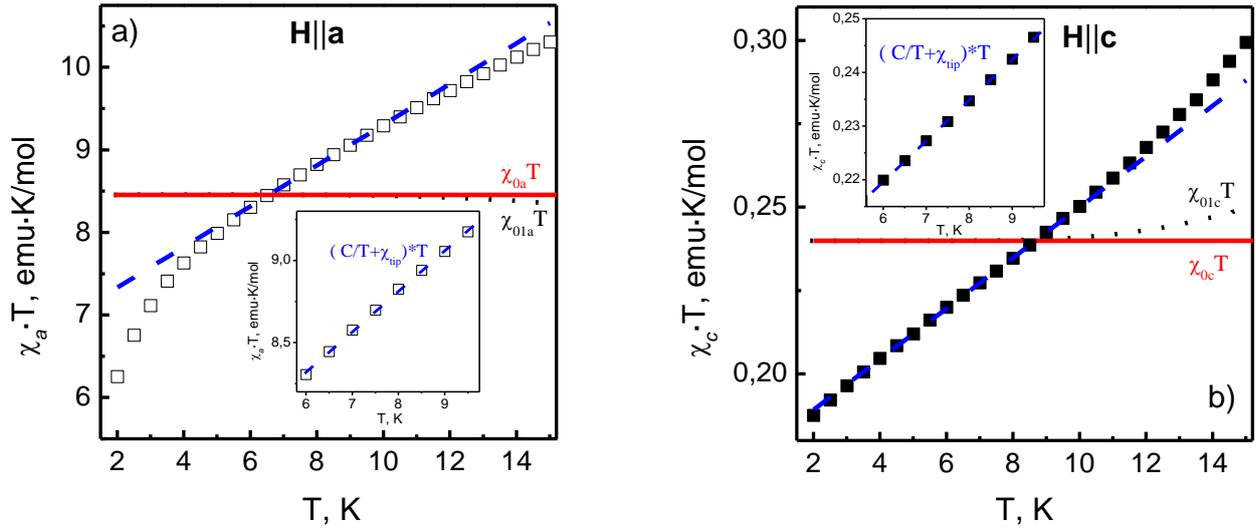

Fig. 2. Temperature dependence of the product $\chi T$ in ErAl$_3$(BO$_3$)$_4$ single crystal for two orientations $\mathbf{H}\parallel\mathbf{a}$ (a) and $\mathbf{H}\parallel\mathbf{c}$ (b). Symbols are the experimental data. Solid lines $\chi_0 T$ represent the calculation results using formula (3) (effective model is "one isolated doublet"). Dotted curves $\chi_{01}T$ correspond to the calculation results using formula (2) (four-level model is "ground + first excited doublet"). The insets show in detail the temperature interval from 6 to 9.5 K, in which a linear regime of product $\chi T$ (see dashed straight lines in inserts and main panels) for two magnetic field orientations is observed.

Thus, temperature dependence the product of $\chi T$ for $\mathbf{H}\parallel\mathbf{a}$ displays three temperature intervals: the low temperature interval ($T < 6$ K), a linear regime (6 K $< T <$ 9.5 K) and the high-temperature interval ($T > 9.5$ K). Since for $\mathbf{H}\parallel\mathbf{c}$ the product of $\chi_c T$ does not show noticeable deviations from linearity at low temperatures, here only two temperature ranges can be distinguished: a linear regime (2 K $< T <$ 9.5 K) and the high-temperature interval ($T > 9.5$ K).

As it can be easily understood, for the segments of temperature dependence of $\chi T$ having a linear regime, the magnetic susceptibility can be described by the following simple expression:

$$\chi(T) = C/T + \chi_{tip} , \quad (1)$$

where $C$ is Curie constant for the $Er^{3+}$ ions and $\chi_{tip}$ is the temperature independent contribution to magnetic susceptibility. For this case magnetic susceptibility $\chi(T)$ can be represented as a sum of two parts: simple Curie law plus some constant value. The estimated values of the Curie constant $C$ and the temperature independent contribution $\chi_{tip}$ were determined by using the values of cutoffs and slopes for dashed straight lines in the temperature range of 6 - 9.5 K (see inserts in Fig. 2). The obtained values for $C$ and $\chi_{tip}$ for $\mathbf{H} \parallel \mathbf{a}$ and $\mathbf{H} \parallel \mathbf{c}$ are summarized in Table 1.

|  | $C$, emu·K/mol | $\chi_{tip}$, emu/mol | $\theta$, K |
|---|---|---|---|
| $\mathbf{H} \parallel \mathbf{a}$ [exp. (1)] | 6.84 | 0.25 | - |
| $\mathbf{H} \parallel \mathbf{a}$ [exp. (4)] | 7.50 | 0.21 | - 0.45 |
| $\mathbf{H} \parallel \mathbf{c}$ [exp. (1)] | 0.17 | $7.60 \cdot 10^{-3}$ | - |

Table 1. Curie constants $C$, temperature independent part $\chi_{tip}$, and Curie-Weiss temperature $\theta$ obtained for $ErAl_3(BO_3)_4$ using two models (see expressions (1) and (4)).

The first term in expression (1) is the Curie law for an ensemble of non-interacting magnetic $Er^{3+}$ ions. The second term is temperature independent part with positive sign, and for two orientations $\mathbf{H} \parallel \mathbf{a}$ and $\mathbf{H} \parallel \mathbf{c}$ the values of $\chi_{tip}$ are quite different. The temperature independent contribution can be related with the different physical nature: experimental systematic shift, diamagnetism, Van Vleck paramagnetism and etc. As a rule, Van Vleck susceptibility exceeds the diamagnetic one at least by an order of magnitude. Therefore, the diamagnetic susceptibility together with possible experimental error was neglected.

It is well known, that the Van Vleck paramagnetism comes from the second order corrections to the Zeeman interaction. This contribution is positive and temperature-independent. The manifestation of the Van Vleck paramagnetism on the temperature dependence of magnetic susceptibility was observed in a crystal $YAlO_3$ doped with the $Er^{3+}$ ions [12]. Thus, a significant Van Vleck contribution to the magnetic susceptibility of $ErAl_3(BO_3)_4$ seems quite possible.

To identify the reasons of deviation of the experimental dependences of susceptibility from formula (1), let us consider at first a simpler case of $\mathbf{H} \parallel \mathbf{c}$. As was already stated above, such deviation is observed only in the range of relatively high temperatures of over 9.5 K. Therefore, it is absolutely natural to consider the possibility of potential influence of the excited levels of the ground multiplet on the susceptibility. In the investigated temperature range, it will be reasonable to take into account only the first of these levels. Then the expression for the magnetic susceptibility of a mole of non-interacting $Er^{3+}$ ions will be as follows:

$$\chi_{01} = \frac{N_A \mu_B}{2H} \frac{g_0(\exp(\frac{g_0\mu_B H}{2kT}) - \exp(-\frac{g_0\mu_B H}{2kT})) + g_1 \exp(-\frac{E_1}{T})(\exp(\frac{g_1\mu_B H}{2kT}) - \exp(-\frac{g_1\mu_B H}{2kT}))}{\exp(\frac{g_0\mu_B H}{2kT}) + \exp(-\frac{g_0\mu_B H}{2kT}) + \exp(-\frac{E_1}{T})(\exp(\frac{g_1\mu_B H}{2kT}) + \exp(-\frac{g_1\mu_B H}{2kT}))} \quad , \quad (2)$$

where $N_A$ is the Avogadro's number, $E_1 = 66$ K (~ 46 cm$^{-1}$) is the energy distance between the ground and the first excited doublets of the $^4I_{15/2}$ multiplet and $g_1$ - is g-factor of the excited doublet. If the first excited doublet is not taken into account, then in low magnetic field $g_0\mu_B H << kT$ the expression (2) can be written as follows:

$$\chi_0 = N_A g_0^2 \mu_B^2 / 4kT \quad . \quad (3)$$

According to the EPR data [9], the value of g-factor of the ground doublet for the field orientation along c-axis is $g_{0c}$=1.6. Due to the absence of any data about g-factor of the first excited doublet in $ErAl_3(BO_3)_4$, the estimated value of $g_{1c}$=3.3 for **H**∥**c** can be used from work [13], where the $YAlO_3$ crystal was doped by the $Er^{3+}$ ions and it has the similar energy splitting $E_1$ between two lowest doublets. The calculation result for the product $\chi_{01c}T$ using formula (2) is shown as dotted line in Fig. 2 (b). To compare, the calculation result for the product $\chi_{0c}T$ using formula (3) is shown as a solid line in the same figure. One can easily noted that noticeable departure of the product $\chi_{01c}T$ from $\chi_{0c}T$ occurs only above 9.5 K. For the same temperature range ($T > 9.5$ K) the departure of experimental data from the linear dependence is observed (see details in Fig.2(b)). Thus, taking into account of the contribution from the first excited doublet leads to the increasing value of $\chi T$ for **H**∥**c** above 9.5 K.

In Fig. 1 (b) the dashed line is plotted by using expression (1) $\chi(T) = C/T + \chi_{tip}$ with the following parameters $C_c = 0.17$ emu·K/mol and $\chi_{tip,c} = 7.6 \cdot 10^{-3}$ emu/mol. As one can see in Fig. 1 (b), by using two parameters, $C$ and $\chi_{tip}$, only we can well describe the experimental data for **H**∥**c** for $ErAl_3(BO_3)_4$ in the whole temperature range from 2 to 15 K.

Similar algorithm of consideration and calculations has been applied for analyzing the field orientation **H**∥**a**. In the frame of four-level model ("ground + first excited doublet") magnetic susceptibility has been calculated using expression (2) with the following values of g-factors corresponding for the a-axis: $g_{0a} = 9.5$ [9], $g_{1a} = 0$ [13]. The result of calculation for the four-level model (product $\chi_{01a}T$) is shown as a dotted line in Fig. 2 (a). In the figure the solid line (marked as $\chi_{0a}T$) is the calculation result for effective model of "one isolated doublet" that corresponds to the expression (3). Taking into account the presence of the first excited doublet leads to the small reducing value of $\chi T$ for $\mathbf{H} \parallel \mathbf{a}$ with increasing temperature above 9.5 K (compare $\chi_{01a}T$ and $\chi_{0a}T$ curves in Fig. 2 (a)).

Below 6 K the low-temperature magnetic properties of $ErAl_3(BO_3)_4$ are also of a range of scientific interest, which is due to the occurrence of possible spin-spin correlations in magnetic system. As you can see in Fig. 2 (a), for $T < 6$ K ($\mathbf{H} \parallel \mathbf{a}$) the increasing deviation of experimental data from the calculated straight line (see dashed line) for decreasing temperature is observed. Such magnetic behavior may indicate the presence of the nonzero interactions between the $Er^{3+}$ ions. In the same time, for similar temperature interval (2 – 6 K) a significant growth of the magnetic contribution to the heat capacity, additional to the phonon contribution, was detected in [10]. This is another proof of occurrence of spin correlations in $ErAl_3(BO_3)_4$. Thus, the observed departure of the experimental data from the linear dependence below 6 K can be considered as a development of a short-range magnetic order in $ErAl_3(BO_3)_4$. Since the experimental data are shifted downward from the expected linear dependence for a system of non-interacting ions, one can conclude that the magnetic interactions between the moments of the $Er^{3+}$ ions have an *antiferromagnetic* character in the studied crystal.

Taking into account the presence of antiferromagnetic interaction between the spins of the $Er^{3+}$ ions in $ErAl_3(BO_3)_4$ leads to the following expression for temperature dependence of magnetic susceptibility:

$$\chi(T) = C/(T-\theta) + \chi_{tip} , \qquad (4)$$

where $\theta$ is the Curie-Weiss temperature, which is negative for an antiferromagnetic interaction. For this model the temperature dependence of magnetic susceptibility $\chi(T)$ can be represented now as a sum of two parts: simple Curie-Weiss law (ensemble of weakly interacting ions) plus some constant value. Therefore, the analysis of the experimental curve in Fig. 2 (a) can be only of qualitative nature, while $C_a$ and $\chi_a$ values cannot be accurate, as these values were obtained by using the expression excluding spin-spin interactions. On the other hand, distinct deviation from the linear temperature dependence $\chi_a T$ at 6 K $< T <$ 9.5 K is not observed (see insert to Fig. 2 (a)). Thus, magnetic susceptibility (as well as the heat capacity [10]) does not reveal the influence of spin correlations above 6 K.

In order to estimate three new parameters, such as a Curie-Weiss temperature $\theta$, a Curie constant $C$ and temperature independent contribution $\chi_{tip}$, we attempted to obtain the best agreement between the expression (4) and the experimental data for $\mathbf{H} \parallel \mathbf{a}$. The solid line in Fig. 1 (a) corresponds to the best fit calculation result with the following parameters: $\theta = -0.45$ K, $C_a = 7.50$ emu·K/mol, $\chi_{tip,a} = 0.21$ emu/mol. Small difference between the experimental data and the calculated curve may be associated with the presence of critical phenomena manifestation during approaching magnetic order temperature, rather than with measurement or parameter determination inaccuracy. From this point of view, further investigation of the ErAl$_3$(BO$_3$)$_4$ single crystal at ultralow temperatures below 2 K where the magnetic ordering is desired. For visual illustration of the influence of the spin-spin interactions on the magnetic susceptibility of ErAl$_3$(BO$_3$)$_4$, the calculated dependence (4) with θ = 0 K (without interaction) is presented as dashed line in Fig. 1 (a).

It should be noted that there is a difference in the estimates of the Curie constant $C$ obtained from the analysis of the results of magnetic measurements and the EPR data. For example, the value of the Curie constant can be obtained using the experimental spectroscopic characteristics (effective splitting value) of the main doublet of a magnetic ion. For a system of non-interacting magnetic Er$^{3+}$ ions, this constant can be calculated with the standard expression $C = N_A g_0^2 \mu_B^2 S(S+1)/3k$, where $g_0$ is the $g$-factor of spectroscopic splitting for particle with effective spin $S=1/2$. For **H** ∥ **a** the EPR data reveal $g_0 = 9.5$ and consequently the estimation of Curie constant is equal $C_a = 8.47$ emu·K/mol. When analyzing the magnetic data, it can be seen that the modified Curie-Weiss law $\chi(T) = C/(T-\theta)+\chi_{tip}$ describes well the temperature dependence of the magnetic susceptibility of the ErAl$_3$(BO$_3$)$_4$ single crystal (see Fig. 1(a)), with an effective parameter equal is $C_a = 7.50$ emu·K/mol, that is by 10 % less when compared to the EPR estimation. Such discrepancy of the Curie constants cannot be due to an inaccuracy of orientation of external magnetic field, since in this case the tilting angle between $H$ and the $a$-axis should be about 20 degrees.

One of the simplest explanations for the discrepancy between the same estimated parameters obtained in the result of magnetic and resonance measurements could be a relatively small decrease (a deficit of no more than 10 %) of the concentration of magnetic centers of the erbium ions in the sample. The magnitude of the molar magnetic susceptibility and, accordingly, the effective magnetic parameters of studied system compare with EPR data can be lowered after the normalization procedure. On the other hand, a small changing in the concentration of magnetic centers in a crystal can also be associated with the presence of some impurity amount, for example, like bismuth and molybdenum, which substitute the Er$^{3+}$ ions in the process of the crystal's growth. The concentration of such impurities in real crystals can be significant [14].

Previously in the work [9, 10] it was shown, that the anisotropic nature of spin-spin interactions in ErAl$_3$(BO$_3$)$_4$ is related to the occurrence of different non-zero initial splittings in EPR spectrum. Our investigations identified that above 2 K spin correlations do not manifest themselves on magnetic susceptibility along the $c$-axis. However, the spin correlations substantially influence the temperature dependence of $\chi_a T$. Thus, we directly determined that spin-spin interaction in ErAl$_3$(BO$_3$)$_4$ single crystal manifests pronounced anisotropy. This fact is in full agreement with the previously expressed suggestion [10] about prevalence of dipole-dipole interaction contribution to the spin-spin correlations between the $Er^{3+}$ ions. It is known, that the value of the dipole-dipole energy of interaction between the neighboring ions at low temperatures is $E_{dd} \sim g_0^2/r^3$, where $r$ is a distance between the $Er^{3+}$ ions. Therefore, it is only due to the difference in values of the $g$-factors that the dipole-dipole interaction in the direction of the $a$-axis over 35 times exceeds $E_{dd}$ along the $c$-axis. Unfortunately, there is a lack of accurate data on structural parameters of the $ErAl_3(BO_3)_4$ single crystal at liquid helium temperatures. However, it is known that for such crystals, the shortest distance between the rare-earth ions is in the direction of the $a$-axis [15, 16]. Thus, due to the cubic term in the denominator in expression $E_{dd} \sim g_0^2/r^3$, the dipole-dipole energy $E_{dd}$ in the direction of the $a$-axis will exceed the one along the $c$-axis even more significantly. In effect, this difference can reach almost two orders of magnitude. It is this fact that results in qualitative difference of spin correlations effects on the investigated low temperature magnetic susceptibility, which was identified in this study.

## CONCLUSION

Low temperature magnetic susceptibility of the $ErAl_3(BO_3)_4$ single crystal in the temperature range from 2 to 15 K for two orientations **H**∥**a** and **H**∥**c** have been investigated. It was shown that qualitatively the magnetic behavior can be described in the framework of simple models (Curie or Curie-Weiss laws) taking into account the temperature-independent contribution to magnetic susceptibility. It was found below 6 K the presence of noticeable spin-spin interactions (the influence on magnetic susceptibility) in the direction of external magnetic field along the $a$-axis. It was shown that the effective interaction between the magnetic moments of the $Er^{3+}$ ions has the antiferromagnetic character. The estimated value of Curie-Weiss temperature is $\theta_{AF} \approx -0.45$ K. In our opinion, the effective magnetic interactions between the $Er^{3+}$ ions in $ErAl_3(BO_3)_4$ are mainly of a dipole-dipole nature.

## ACKNOWLEDGMENTS

This work was supported partially by the National Science Centre, Poland, under project No. 2018/31/B/ST3/03289.